# FFREE: a Fresnel-FRee Experiment for EPICS, the EELT planets imager

Jacopo Antichi[a], Christophe Vérinaud[a], Olivier Preis[a], Alain Delboulbé[a], Gérard Zins[a], Patrick Rabou[a], Jean-Luc Beuzit[a], Sarah Dandy[b], Jean-François Sauvage[b], Thierry Fusco[b], Emmanuel Aller-Carpentier[c], Markus Kasper[c], Norbert Hubin[c]

[a]Laboratoire d'Astrophysique de l'Observatoire de Grenoble, 414 Rue de la piscine F-38041 France;
[b]Office National d'Etudes et Recherches Aérospatiales, 29 Avenue Leclerc, F-92332 France;
[c]European Southern Observatory, Karl Schwarzschild Str. 2, D-85748, Germany

## ABSTRACT

The purpose of FFREE - the new optical bench devoted to experiments on high-contrast imaging at LAOG - consists in the validation of algorithms based on off-line calibration techniques and adaptive optics (AO) respectively for the wavefront measurement and its compensation. The aim is the rejection of the static speckles pattern arising in a focal plane after a diffraction suppression system (based on apodization or coronagraphy) by wavefront pre-compensation. To this aim, FFREE has been optimized to minimize Fresnel propagation over a large near infrared (NIR) bandwidth in a way allowing efficient rejection up to the AO control radius, it stands then as a demonstrator for the future implementation of the optics that will be common to the scientific instrumentation installed on EPICS.

## 1. INTRODUCTION

Detection of exoplanets implies to measure extremely weak signals, usually below intensity perturbations in the Point Spread Function (PSF) caused by static aberrations. Hence, two AO correction stages of aberrations need to be foreseen. The first aimed at the correction of the atmospheric (dynamic) turbulence, the second aimed at correction of the residual static aberrations due to non common path between the sensor and the instrument (i.e. residual optical aberration, optics alignments, optics polishing, etc.) before the beam reaches the scientific focal-planes. The definition and design of the second stage needs a laboratory experiment where optical design and mechanical assembling are the result of AO-simulations including the impact of the physical propagation of the electric field (i.e. Fresnel propagation) collected by the telescope pupil. Only within such an integrated setup it is possible to achieve this cancellation purpose exploiting off-line calibration techniques, resumed here as *phase diversity* techniques, for the wavefront measurement and the AO control for the wavefront pre-compensation.

The scope of this experiment is to demonstrate the use of these techniques in the framework the EPICS common path optical design and its mechanical implementation.

## 2. RATIONALE

The leading argument for an optical concept free from the effects of Fresnel propagation [1] and contemporary exploiting AO to measure and compensate the wave front through phase diversity [2] lies on the imaging properties of the adopted deformable mirror (DM). The starting point is then that the minimum Talbot length ($z_T$), defined as:

$$z_T \equiv \frac{2 \cdot \Lambda^2}{\lambda}, \qquad (2\text{-}1)$$

($\Lambda$ being a spatial period and $\lambda$ a wavelength) can be fixed by the DM pitch size (P) exploiting the following relation:

$$z_T = \frac{2 \cdot (2 \cdot P)^2}{\lambda}. \qquad (2\text{-}2)$$

The DM adopted in FFREE has an actuator pitch equal to 340 micron; the FFREE Talbot length is then: 600 mm.

More in details, any single periodic component of a static aberration pattern projected onto the DM is working - in terms of optical propagation - like a diffractive grating pattern [3]. Thus, after the reflection onto the DM this pattern propagates according to the Fresnel prescription, i.e. reproducing itself at distances equal to its corresponding Talbot length [4]. Moreover, the fact that it is possible to fix a minimum Talbot length in terms of the DM pitch lies on the Shannon sampling theorem. In fact, while single components of a static aberrations pattern with spatial period $\Lambda \geq 2 \cdot P$ will be sampled according to Nyquist, any other component with higher spatial frequency will be aliased. Hence a Fresnel-free zone in the field can be fixed by a first principle, i.e. by making use of the concept of AO control radius.

Considering all these points, it turns that the selected DM on FFREE (32 x 32 actuators, D = 10.88 mm) is well suited for our purposes in the angular scale of an 8 meter pupil. In fact, being P = 340 micron, the Nyquist spatial period returns to be $\Lambda_{NYQUIST}$ = 680 micron. Fixing the angular resolution of an 8 meter pupil as the ones of the VLT and the monochromatic wavelength adopted in the experiment ($\lambda$ = 1.548 micron, $\lambda/2D_{VLT}$ = 0.02 arcsec) it is possible then to calculate the angular position corresponding to the Nyquist spatial period as projected on sky ($\Lambda_{VLT}$):

$$\frac{\lambda}{\Lambda_{VLT}} = \frac{\lambda}{2D_{VLT}} \cdot \frac{2D_{VLT}}{\Lambda_{VLT}} = \frac{\lambda}{2D_{VLT}} \cdot \frac{2D}{\Lambda_{NYQUIST}}. \tag{2-3}$$

By eq. (2-3) it turns that, exploiting the whole surface of our DM, it is possible to obtain a Fresnel-free optical design within $\lambda/\Lambda_{VLT} = \pm 0.64$ arcsec, or – viceversa – fixing a priori the needed Fresnel-free field to be $\lambda/\Lambda_{VLT} = \pm 0.48$ arcsec, it is possible to select the target pupil size as projected onto the DM surface: 8.16 mm; 0.48 arcsec ($\theta_{FFREE}$) is then the field of view of our experiment.

These calculations show the feasibility of a Fresnel-free optical system where Fraunhofer propagation is allowed to be truly representative of the effective complex amplitude optical propagation.

## 3. OPTICAL CONCEPT AND SIMULATIONS

The choice adopted to make a design free from Fresnel aberrations is to exploit pupil magnification as ruler for the minimal Talbot length allowed in the beam propagation. The tenet is properly that any out-of-pupil optics (N) – apart of 2 specific custom optics which lie in front of the DM – can be designed to have distance ($d_N$) wrt. their closest pupil plane verifying the condition:

$$d_N \ll z_T^N, \tag{2-4}$$

where $z_T^N$ is the minimum Talbot length resulting from the Nyquist spatial period ($\Lambda_{NYQUIST}$) as projected by the N-th out-of-pupil optics onto the pupil plane the closest wrt. this specific optics within FFREE:

$$z_T^N \equiv \frac{2\Lambda_N^2}{\lambda} = \frac{2(m_N \cdot \Lambda_{NYQUIST})^2}{\lambda}. \tag{2-5}$$

The selected magnification is m = 4. This number comes out to be a good compromise among different factors: the fixed area of the adopted optical bench, the availability of optics having standard effective focal length (EFFL) and standard aperture sizes compared to the minimum Talbot length equals (600 mm). This choice allows a re-scaling of the Talbot length equal to 9.6 m, thus allowing to any out-of-pupil optics adopted in the design, and re-imaging the telescope pupil, to not introduce Frensel effects onto the beam.

In conclusion, Fraunhofer propagation can easily return to be a good approximation of the electric field propagation, by selecting a suited and unique optical magnification ($m_N = m$) between the DM pupil and any other imaged pupil plane in the system.

### 3.1 OPTICAL PROPAGATION: FRAUNHOFER CASE

In our design optimization the selected DM pupil size is D = 8.16 mm and the adopted wavelength is the one corresponding to a SLED monochromatic NIR source with $\lambda$ = 1.548 micron, in use at the lab. The adopted simulation code is PROPER® [5] which computes the physical propagation between 2 conjugated optical planes implementing the

Fresnel approximation of the Huygens-Fresnel principle [6] for the propagation of the electric field complex amplitude. The adopted ideal optics is a lens with full aperture equals 12.7 mm and EFFL = 75 mm. The Fresnel number (FN) corresponding to this ideal optics is FN = 573 (i.e. >>1) assuring that Fraunhofer propagation is an excellent approximation of the ideal (no phase aberrations in) physical propagation.

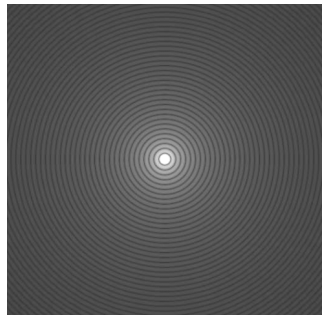

Figure 1: Perfect PSF corresponding to the selected DM pupil (8.16 mm), the images edges correspond to ± 1 arcsec.

The adopted diffraction suppression system in FFREE is pupil apodization with or without the use of a Lyot Stop. The PSF apodized amplitude profile is obtained with the Gerschberg & Saxton iterative method [7]: the amplitude apodized map varies until the requested contrast gain is achieved within $\theta_{FFREE}$.

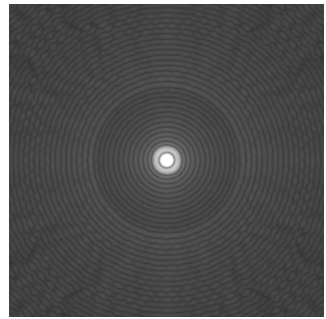

Figure 2: Perfect PSF corresponding to the selected DM pupil after amplitude apodization.

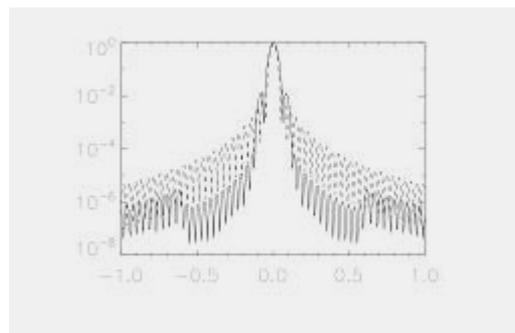

Figure 3: Normalized PSF profiles with (continues line) and without (dashed line) pupil apodization at the convergence of the Gerschberg-Saxton loop: the goal in contrast gain ($10^{-2}$ at ± 0.25 arcsec) is achieved and the coronagraphic PSF is well rejected within $\theta_{FFREE}$ (horizontal axis is in arcsec).

## 3.2 OPTICAL PROPAGATION: FRESNEL CASE

In the case a true out-of-pupil optics is considered to make the DM-pupil to focus conjugation, simulations should start by fixing both the power spectral density (PSD) of the static aberrations pattern corresponding to this specific optics as its root means square (RMS) surface error ($\sigma_{SE}$) and its refraction index (n). According to optical theory, in the spatial frequency space (f), 1-dimensional PSD and $\sigma_{SE}$ are related as follows:

$$\text{PSD}(f) = C \cdot f^{-\alpha}, \tag{2-6}$$

$$\sigma_{WFE}^2 = \int_{1/D}^{\infty} df\, 2\pi\, f \cdot \text{PSD}(f). \tag{2-7}$$

Once the refraction index (n) is fixed, it is possible to link the wavefront error with the surface quality of the optics adopted in the setup. All the optics adopted in FFREE are achromatic doublets average refraction index n = 1.7 with a very small variation between visible and NIR values (< 1%). For a the single optics is then possible to translate the requested wavefront error in the (observable) surface quality considering all the air-glass surfaces of this system (i.e. 2 surfaces) and neglecting the glass-glass surface inside the doublet itself. The relation between wavefront error and the surface error proper to the single optical surface (considered constant over the whole amount of optical surface of the chosen setup) is:

$$\sigma_{WFE}^2 := \sqrt{\sum_{i=1}^{n=2} \sigma_{WFE}^2(i)} = 2 \cdot (n-1)^2 \cdot \sigma_{SE}^2, \tag{2-8}$$

or

$$\sigma_{WFE} = \sigma_{SE} \cdot \sqrt{2} \cdot (n-1) = \sigma_{SE} \cdot \sqrt{2} \cdot 0.7 \approx \sigma_{SE}. \tag{2-9}$$

Eq. (2-9) indicates that it is possible to vary the PSD amplitude (C) and the PSD slope (n) up to a fixed boundary (equals to the inverse of DM pupil image i.e. $D^{-1}$), in order to regulate the optical quality to the needs requested in the simulations. The adopted PSD slope is the "standard" value $\alpha = -2$, while the PSD amplitude C varies between 2 extreme cases: the first giving as output RMS wavefront error $\sigma_{WFE}$ = 16 nm (or an equivalent RMS surface error $\sigma_{SE}$ = 16 nm i.e. $\lambda$/40 at 0.6328 micron) and the last giving as output RMS wavefront error $\sigma_{WFE}$ = 40 nm (or an equivalent RMS surface error $\sigma_{SE}$ = 40 nm i.e. lambda/16 at 0.6328 micron). Finally, as already explained, the adopted EFFL imposed to the out-of-pupil optics allowing the DM-pupil to focus conjugation is fixed to be 75 mm. This value is selected as "skill" compromise between the need to minimize this length wrt. the minimum Talbot length in the bench, and the practical fact that lowering the EFFL of a real optics – ideally placing it a distance zero wrt. to DM – it means to force the optics itself to be in double-pass with the DM itself. This fact – in turn – is not good in terms of static design aberration: double-pass implies to use such optics in some off-axis mode inducing the entire system to acquire an unwanted amount of astigmastism.

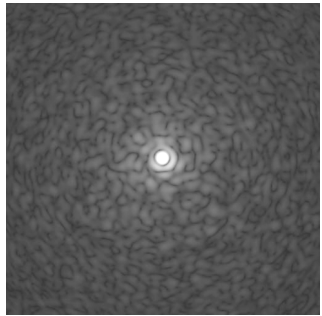

Figure 4: PSF corresponding to the selected DM pupil through an out-of-pupil optics with EFFL = 75 mm and $\sigma_{SE}$ = 40 nm: no AO compensation is applied and speckles induced by Fresnel propagation appear also within $\theta_{FFREE}$.

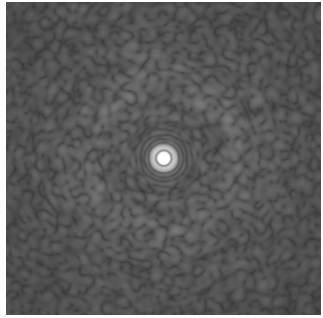

Figure 5: PSF corresponding to the selected size pupil and the same out-of-pupil optics, but with an ideal AO compensation: aliased spatial frequencies in the sampled wavefront are perfectly filtered out. Apodization is inserted in the complex amplitude before fast Fourier transform (FFT) provides the focal image: Fresnel-induced speckles still dominate within $\theta_{FFREE}$.

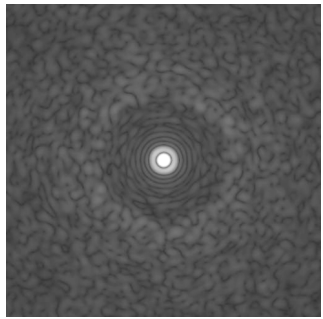

Figure 6: PSF corresponding to the selected size pupil and the same out-of-pupil optics, but with a different surface quality ($\sigma_{SE}$ = 16 nm): this choice allows restoring to its diffraction limit the final image, within $\theta_{FFREE}$.

It important to stress that in this context residual dynamical aberrations and residual high-frequency static aberrations due to the telescope optics are not taken into account and the term *diffraction-limit* refers specifically to the hypothesis that, within a fixed angular boundary, Fraunhofer propagation is truly representative of the effective optical propagation.

### 3.3 AO COMPENSATION WITH PHASE DIVERSITY: SIMULATIONS FIRST RESULTS

The correction of static aberrations requires the use of a sensor able to measure and compensate the wavefront up to the higher spatial frequencies. Phase diversity can be applied to this context as a focal plane wavefront sensor allowing potentially to reach high accuracies because it allows to measure and to compensate for a sound number of modes and because it does not need extra optical devices, avoiding – de facto – non common path static aberrations.

In order to estimate how phase diversity performs the measure and the compensation of the wavefront in the working case of FFREE, a first end-to-end simulations run has been done in 2009 exploiting the so called *pseudo-closed loop* method proposed by [8] with adaptations to work with an apodized pupil both in the image formation as in the phase diversity reconstruction. Detailed sensitivity analysis allowed us to establish the robustness of the algorithm. Future work will includes the Fresnel effect impacts on the selected compensation method.

In the simulations, we consider a 32 x 32 pixel pupil, and a total contribution of static aberrations equals 30 nm RMS, with a PSD low proportional to $f^2$ (f being the spatial frequency coordinate). Images are Nyquist sampled at a wavelength of 950 nm (the cut-on wavelength of the NIR arms as foreseen in EPICS). The simulated DM gets 16 x16 actuators and Gaussian influence functions. All the simulations are performed for 2 cases (non apodized or classical pupil and apodized pupil) both without introducing any noise.

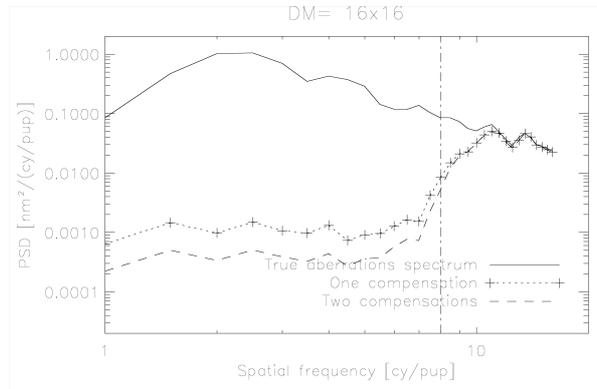

Figure 7: (Non apodized pupil case) estimated static aberrations residual PSD after phase diversity with DM compensation as a function of the number of iterations of the pseudo-closed loop recipe. The decrease of low frequencies just after two compensations validates the selected method. The Nyquist spatial frequency is shown with a vertical line.

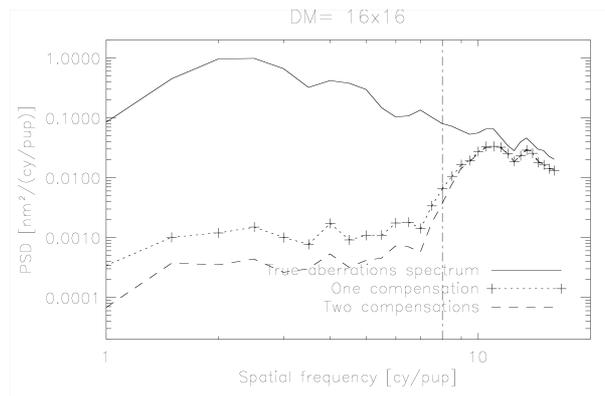

Figure 8: (Apodized pupil case) estimated static aberrations residual PSD after phase diversity with DM compensation as a function of the number of iterations of the pseudo-closed loop recipe. The decrease of low frequencies just after two compensations validates the selected method also in the case of pupil configuration. The Nyquist spatial frequency is shown with a vertical line.

This first run of simulations indicated that the phase diversity measurement as well as the DM compensation allows an adequate correction with small margin between classical and apodized cases. The pseudo-closed loop is then verified to be compliant with the main specification of the FFREE bench, i.e. with the apodization of its pre-coronagraphic pupil.

### 3.4 COHERENT CANCELLATION: MAIN ARGUMENTS

Fresnel propagation is taken into account in order to avoid speckles created by the beam propagation through out-of-pupil optics within a fixed field of view. Inside this, the pre-compensation of a suitable DM should cancel the residual low-order aberrations and allow the system to explore the performances of coronagraphic techniques that should be Fresnel-friendly, i.e. not introducing phase or amplitude errors before the final detector focal plane.

The concept adopted in FFREE makes use only of a pure apodization of the pupil (PAP), dismissing the whole sample of techniques gathered under the name of *Lyot coronagraphs* [9].

In details, the PAP concept arises in FFREE as the result of several basic considerations upon the chromatic non-linearity of real focal plane coronagraphs and the impact of Fresnel propagation of such kind of devices. These are resumed as follows:

- Lyot coronagraphs, i.e. the ones which foresee an amplitude focal mask aimed at the cancellation of the on-axis coherent complex amplitude, introduce ever non linear effects is the resulting chromatic PSF. Adopting $1/\lambda$ as coordinate in the post-coronagraph focus, this means that the chromatic scaling of the PSF is non-linear, hence this represents a source of static speckles which are not caused by physical (Fresnel) propagation of the complex amplitude or by phase residuals in the pupil, not compensated in the AO loop,
- pupil amplitude apodization can be forced to be the unique factor which operates the requested on-axis cancellation, without any further re-imaging of the pupil, that – conventionally – is inserted in the so called *Lyot stop*.

In the real case of FFREE the second item is not fully respected just because it is impossible to deposit such a focal mask directly onto the detector surface. Suited relays optics, conjugating the focal mask focus to the detector focus, should be present then in the design of this demonstrator. The good news is that, within a Fresnel-free optical setup, this last optics does not introduce any further static aberration (a part from residual design aberrations which remain low spatial frequency errors). Differently, what is important is that this optics should oversize the geometrical re-imaged pupil as much as requested to let the FFT to be truly representative of the optical propagation. Within such an optical setup, focal mask focus and detector focus are quasi-stigmatic, and non-linear effects due to the PSF chromaticity are avoided by design.

At contrary, in the case of EPICS it is easy to respect also the second item stated above, just because it is possible to re-image the coronagraphic pupil onto a focal plane at the entrance of the instruments, e.g. a suited integral field spectrograph [3]. Then, the unique effort is to make non-transmissive several lenses of the lenslet array covering the central PSF peak, without any Lyot pupil re-imaging. In this way, coronagraphy is reduced really to a pure amplitude pupil apodization.

## 4. OPTICAL DESIGN

FFREE is an imaging system conceived to provide diffraction limited images in the wavelength range 1.548-1.620 micron. The system is optimized at 1.548 micron, to provide a Nyquist sampled image onto a 320 x 256 infrared detector with pixel size equals 30 microns on a field of view equal to $\theta_{FOV} = \pm 2.3$ arcsec, taking as a reference an 8 meter pupil.

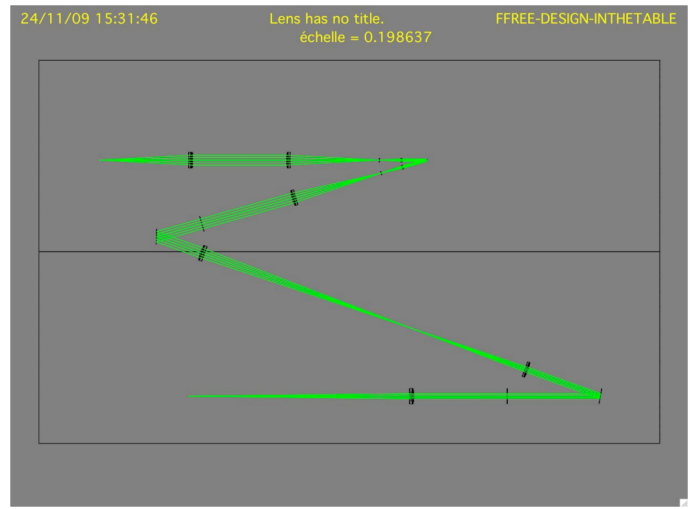

Figure 9: FFREE optical design as implemented in the LAOG bench.

In details, FFREE is an optical setup exploiting a single DM in pseudo-closed loop for the phase measurement and compensation, which is assembled with standard optics except for 2 optics (L3 and L4) named twin lenses (TWL) which are requested to have custom optical quality due to a wise compromise between the request to minimize their EFFL and the one to avoid off-axis configurations in front of the DM. The pupil is imaged 4 times assuring exit-pupil telecentricity and an optical magnification of 4 between the aperture-stop (L0) and the DM pupil, and between the DM pupil and the coronagraphic pupil (in between L4 and M2). A focal mask in the intermediate focus (between L5 and L6) provides the cancellation of the central PSF peak, as requested in coronagraphy. A relays optics (after L6) allows to re-image the focal plane with the correct scale assuring the Nyquist sampling onto the final image plane. This design is thought to be a good compromise between the constraints imposed by Fresnel propagation and a clever selection of catalogue optics. All the catalogue optics are Throlab® achromatic doublets AR-coated in the NIR, made with the same pair of glasses: BAFN10 and SFL6 (a part of L7 made of SF10 and SFL6) and having an average refraction index of 1.733 at λ = 0.6328 micron and 1.705 at λ = 1.548 micron. Only L3 and L4 are custom optics by SILO® made with the same pair of glasses and the same AR-coating. The surface quality of all the catalogue optics is λ/4 peak to valley (PTV) at 0.6328 micron, while the surface quality of the 2 custom optics is λ/10 PTV at 0.6328 micron.

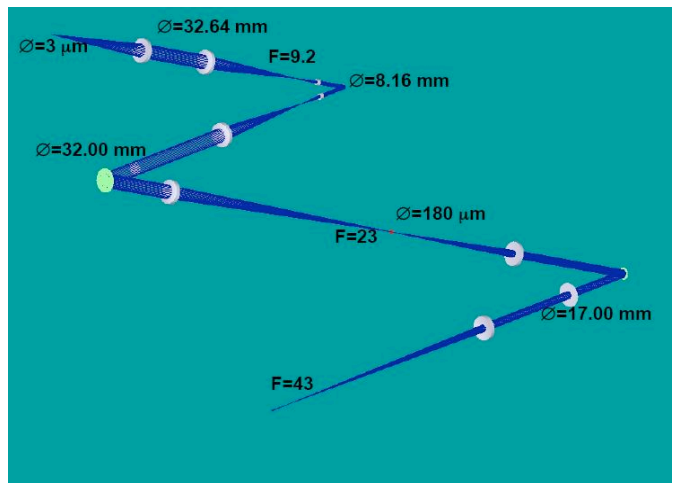

Figure 10: FFREE design main optical parameters when working at λ= 1.548 micron.

## 5. BENCH IMPLEMENTATION

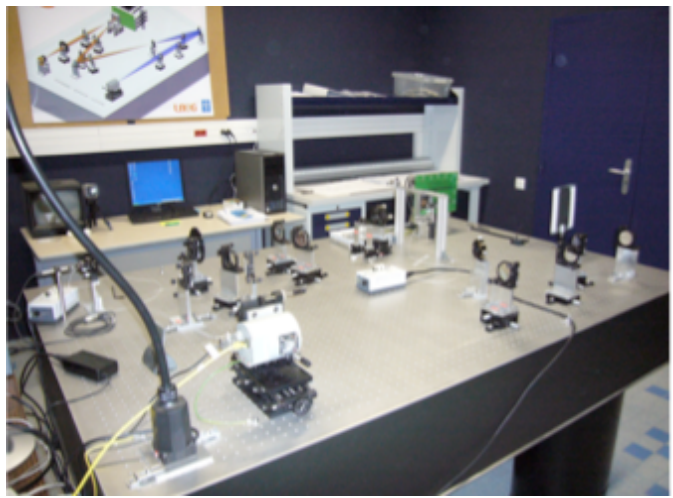

Figure 11: FFREE as implemented at the LOAG bench: source and NIR detector are visible on the left, DM electronics on the right (green color) while the real time controller (RTC) is under the bench.

# 6. CHARCTERIZATION OF THE MAIN SYSTEM COMPONENTS

## 6.1 LIGHT SOURCES

The adopted light sources are 2: a broadband (600 nm) SLED with wavelength $\lambda= 1.548$ micron and a monochromatic LASER with wavelength $\lambda= 0.6328$ micron. The SLED source is coupled with a mono-mode fiber ($NA_{FIBER} = 0.13$) and the LASER source is coupled with a mono-mode fiber ($NA_{FIBER} = 0.12$) too. The LASER source is adopted only during alignments while the SLED source is the one adopted during the experiment. In general, a mono-mode fiber produces as output a coherent beam having a Gaussian intensity. Thus, 3 micron wide pinhole provided by MELLES & GRIOT® is placed downstream the fiber output and assures a beam uniformity > 90% when working with the NIR source.

## 6.2 TWL

According the FFREE optical design, L2 and L3 have been requested to get surface quality to $\lambda/40$ RMS @ $\lambda = 0.6328$ micron, or – in fringe unit – equals 0.05, or – in physical unit equals 16 nm.

Table 1: Interferometric surface quality test results of the FFREE custom optics.

| 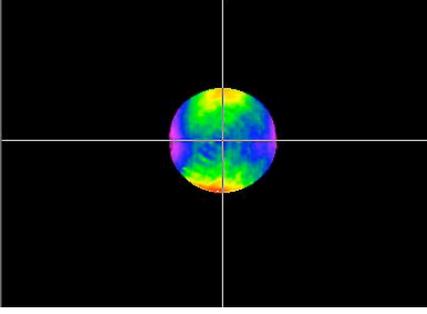 | 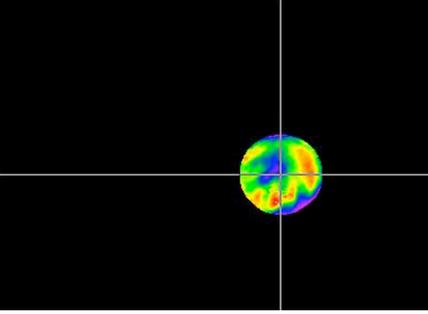 |
|---|---|
| a) L2 left side; $\sigma_{SE} = 0.0163 < 0.0250$: in spec. | b) L2 right side; $\sigma_{SE} = 0.0180 < 0.0250$: in spec. |
| 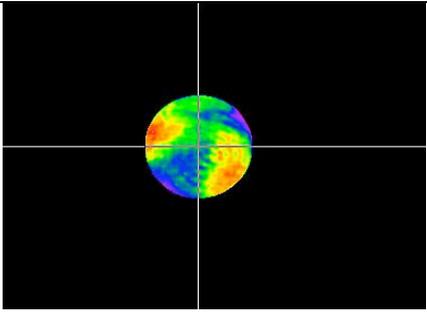 | 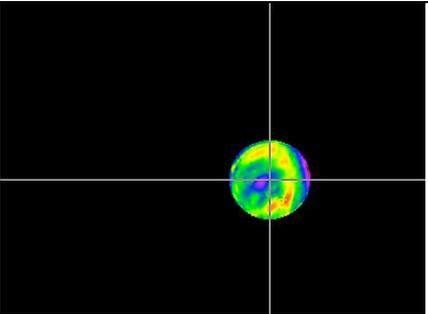 |
| c) L4 left side; $\sigma_{SE} = 0.0131 < 0.0250$: in spec. | d) L4 right side; $\sigma_{SE} = 0.0143 < 0.025$: in spec. |

## 6.3 DM

The adopted DM is BMM®. The default virtual mapping has been discovered to be partially wrong once the DM has been installed on the FFREE bench. The novel virtual mapping done at LAOG matched with the real one, and allowed us to make inspection of the mirror and the test the RTC VLT-software (SPARTA) on to this DM.

The 4 defaults inactive actuators (connections to the ground and tip/tilt commands) have been found as shown in the Vendor virtual map. Beyond these ones, we found 3 dead actuators (#46, #624, #956) while the dead actuator (# 899) – as indicated in the Vendor virtual map − is actually ok.

Table 2: Virtual vs. real maps of the adopted DM (highlighted square indicates the true size of the FFREE DM pupil).

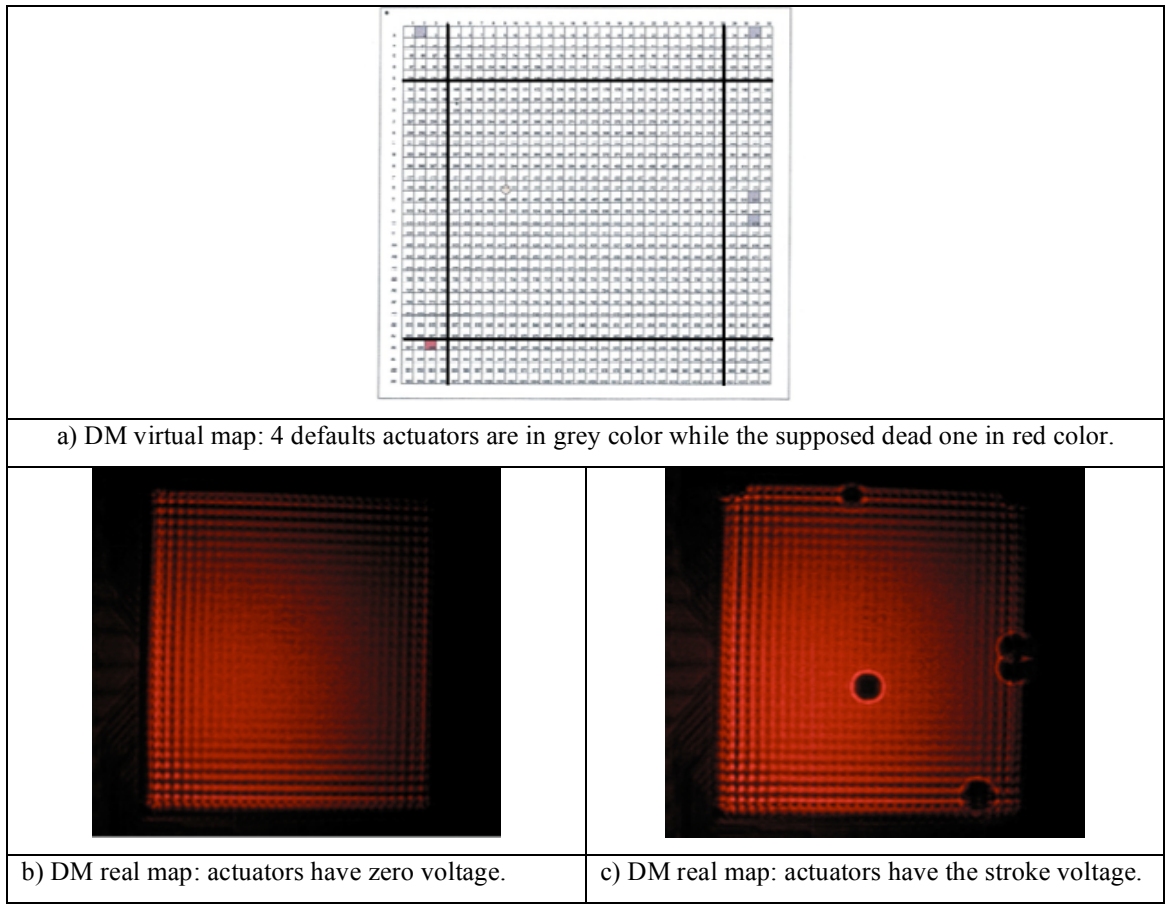

| a) DM virtual map: 4 defaults actuators are in grey color while the supposed dead one in red color. ||
|---|---|
| b) DM real map: actuators have zero voltage. | c) DM real map: actuators have the stroke voltage. |

### 6.4 PUPIL APODIZERS AND FOCAL MASKS

The pupil apodizers and focal masks are provided by POI® and obtained by the use of the halftoning technique, the process of presenting a continuous image through use of dots, by which the electric field amplitude transmission is controlled following the prescription of simulations with confirmed reliability [10].

The apodizers are 2 amplitude masks (circular and central-symmetric) placed in a suited pupil plane of the FFREE bench (between L4 and M2 according Figure 10) having 2 different apodization profiles that follow 2 distinct prescriptions for the central pupil obscuration ratio (15% and 30%) respectively corresponding to the linear obstruction ratio of the VLT and EELT Nasmyth foci.

The focal masks are 2 amplitude masks (circular and central-symmetric) placed in a suited focal plane of the FFREE bench (between L5 & L6 according Figure 10) having with the same diameter (180 micron corresponds to ~ 4.7·$\lambda$/D [11] in the scale of the FFREE coronagraphic focus) and same transmission profile but distinct NIR optical density (the first between 3 and 4, the second greater than 6) in order to test 2 different coronagraphic attenuation regimes.

Table 3: Pupil apodizers normalized ideal vs. real transmission.

| 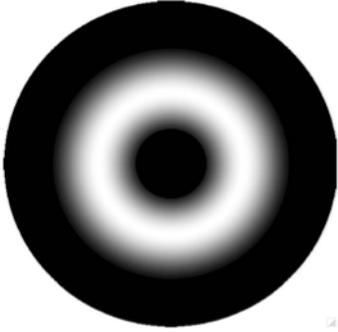 | 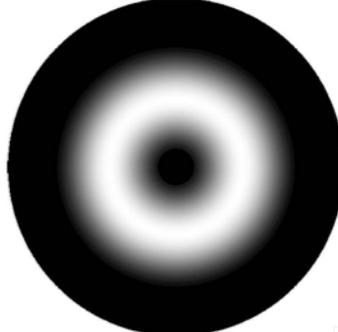 |
|---|---|
| a) PAP ideal transmission for an EELT-like obstruction. | b) PAP ideal transmission for a VLT-like obstruction. |
| 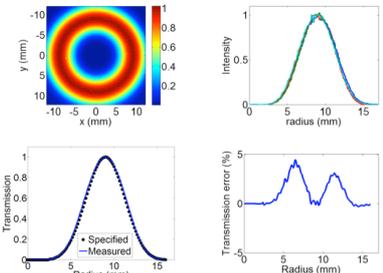 | 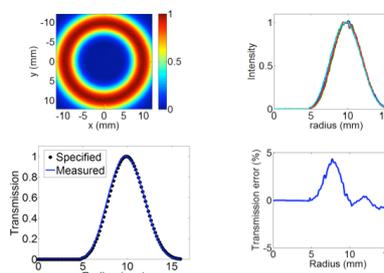 |
| c) PAP measured transmission for an EELT-like obstruction. | d) PAP measured transmission for a VLT-like obstruction. |

## 6.5 NIR DETECTOR

The NIR detector selected for FFREE is the NIR-300PGE provided by VDS® having 256 x 320 pixels and pixel size 30 micron. Its main property lies in the technology that provides the detector to be sensible to NIR radiation with an integrated Peltier cooling system. This in turn gets the cons of a quite high read-out-noise (RON < 400 electrons according to the datasheet) which is a factor of 25 higher than the one of scientific cooled NIR detectors like HAWAII-2RG. However, as shown in Table 4, its flat-field quality is compelling with such kind of detectors.

Table 4: Characteristics of the FFREE NIR detector for a data integration time of 20 ms.

| 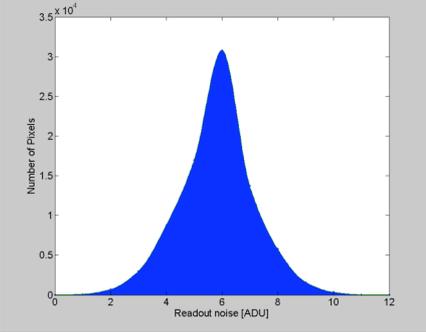 | 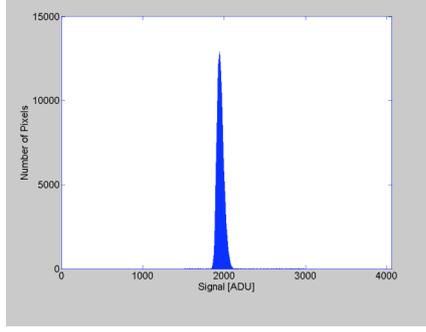 |
|---|---|
| a) RON histogram for a dark exposure. | b) Flat field histogram for a uniform exposure. |

## 7. CONCLUSIONS

The goal of the FFREE experiment for the Framework Program 7 preparatory phase in Europe is to validate experimentally active speckles rejection with the solution chosen for the EPICS baseline in the NIR: PAP, namely diffraction rejection by apodization only [12]. Thanks to a very high achromaticity of the apodizer and the bench itself, speckles rejection on the whole AO control radius and broad spectral band is expected.

After validation of active correction, the FFREE bench will be used to validate high accuracy speckles spectral calibration by post-processing with an integral field spectrograph (IFS). The goal is to confront Fresnel simulations of common path errors by the experiment.

The foreseen extension of FFREE over mid- 2010-2011 will consist in:
- the development of a system of movable refractive phase screens to introduce known patters due to Fresnel propagation,
- the integration of a tunable LASER in order to record speckles position and intensity in function of wavelength and to confront results to Fresnel models.

Ultimately, this bench will also be able to test advanced coronagraphs and issues related to diffraction-limited IFS, like Fresnel-induced spectrograph's non-linearity.

This work is partially funded by the European Union, Framework Program 7 "European Extremely Large Telescope preparatory phase" under Contract NO INFRA-2.2.1.28 and by the European Southern Observatory, EPICS phase-A study.